\documentstyle[12pt]{article}
\begin{document}
\renewcommand{\thefootnote}{\fnsymbol{footnote}}
\large
\date{ }
\begin{center}
{\Large Constraints on new interactions \\ from neutron scattering
experiments}

\vskip 0.5cm

Yu.N.Pokotilovski\footnote{ E-mail: pokot@nf.jinr.ru}

\vskip 0.5cm
            Joint Institute for Nuclear Research\\
              141980 Dubna, Moscow region, Russia\\
\vskip 0.5cm

\begin{minipage}{130mm}

\vskip 0.5cm

Constraints for the constants of hypothetical Yukawa-type corrections to the
Newtonian gravitational potential are obtained from analysis of neutron
scattering experiments. Restrictions are obtained for the interaction range
between 10$^{-12}$ and 10$^{-7}$ cm, where Casimir force experiments and
atomic force microscopy are not sensitive. Experimental limits are obtained
also for non-electromagnetic inverse power law neutron-nucleus potentials.
Some possibilities are discussed to strengthen these constraints.

\end{minipage}
\end{center}

\vspace{1cm}

\section{Introduction.}
Hypothetical new long-range interactions have been discussed from
different points of view. First, developments in quantum field
theory have lead to the possibility of the existence of number of
light and massless elementary particles \cite{Kim}. Exchange of
such particles between two bodies can reveal itself as an
additional interaction potential of Yukawa or power-type and
result in a deviation of the gravitational force from Newtonian
law. On the other hand the theoretical models have been developed
recently with extra spatial dimensions \cite{Ark} leading to an
additional forces at short distances. The deviations of results of
measurements of gravity forces at macroscopic distances from
calculations based on Newtonian physics can be seen in the
experiments of Galileo-, E\"otv\"os- or Cavendish-type
\cite{Fisch} performed with macro-bodies. At smaller distances
($10^{-7}-10^{-2}$) cm the effect of these forces can be observed
in measurements of the Casimir force between closely placed
macro-bodies (for review see \cite{Rep}) or in the atomic force
microscopy experiments. At even smaller distances such experiments
are not sensitive enough, and high precision particle scattering
experiments may play their role. In view of absence of electric
charge the experiments with neutrons are more sensitive than with
charged particles, electromagnetic effects in scattering of
neutrons by nuclei are generally known and can be accounted for
with high precision. As an example of precision scattering
experiments we present here the analysis of the measured total
neutron scattering cross section by heavy nuclei in the eV-keV
energy range, and consider some possibility to increase the
sensitivity in inferring new interactions from the neutron
scattering experiments.

\section{Non-Newtonian gravity.}
For the potential of the form
\begin{equation} U(r)=\frac{g}{r}e^{-r/\lambda}
\end{equation}
where $g=-\alpha GMm$, $G$ -- the Newtonian gravitational
constant, $M$ and $m$ are the masses of gravitating bodies
(nuclear and neutron), the Born amplitude for the scattering by
homogeneous gravitating ball (nucleus) with radius $R$ is
\begin{equation}
f(q)=\frac{2mg\lambda^{2}}{\hbar^{2}(q^{2}\lambda^{2}+1)}F(q).
\end{equation}
Here the nuclear formfactor is
\begin{equation}
F(q)=\frac{3}{(qR)^{3}}\Bigl[{\mbox{sin}}(qR)-qR{\mbox{cos}}(qR)\Bigr],
\end{equation}
where $R$ is nuclear radius.

Inverse power-($N$+1) gravitation potentials
\begin{equation}
U_{\mbox{extra}}(r)=\frac{g_{N}}{r}\Bigl(\frac{r_{0}}{r}\Bigr)^{N},
\end{equation}
where $g_{N}=-\alpha_{N}GMm$, $r_{0}$ is arbitrary length and may
be taken 1 fm, and $N$=1,2,... appear \cite{Ark} in models with
large extra dimensions for $r<R_{c}$, ($R_{c}$ is the
characteristic compactification radius).

The value of the interaction radius $\lambda$ in the Yukawa-type
potential of the Eq.(1), the power $N$ in the power-law potential
of the Eq.(4), the compactification radius $R_{c}$ of extra
dimensions and the strengths of these interactions $\alpha$ and
$\alpha_{N}$ expressed here as their ratios to the newtonian
interaction, are not predicted by theory and are the subject of
experimental investigation. Restrictions for the Yukawa-type
potential down to 10$^{-8}$ cm range were obtained previously from
experiments with macro-bodies. The bounds obtained in these
experiments on the value $\alpha$ increase from $\sim$1 at the
millimeter distances to $10^{35}$ at the distances $\sim$10$^{-8}$
cm. The summary of these results is shown in the Figure 1 with
corresponding referencies.

\section{Low energy neutron scattering -- basic relations.}
The problem of inferring the admixture of the weak long-range
potentials from neutron cross section data is not trivial. During
the years of attempts of the measurements the values of
neutron-electron and neutron electric polarizability scattering
amplitudes different approaches were used with different success
(for review of some of these methods see \cite{Sam}). The most
precise data for these constants were obtained from the
measurements of the neutron total scattering cross section by
heavy nuclei, the differential cross section generally are
measured with much lower precision. The contribution of the
neutron-electron interaction to the neutron-nucleus scattering
cross section is proportional to the electric charge of the
nucleus, the contribution of the neutron electric polarizability
is proportional to the square of the charge of the nucleus,
therefore the measurements of scattering are performed for heavy
nuclei. The contribution of non-Newtonian gravity should be
proportional to the mass of the nucleus, it means that in this
case the heavy nuclei should give the largest effect. On the other
hand analysis of neutron scattering by heavy nuclei presents the
most serious difficulties in view of great effect of resonances in
neutron scattering. For light nuclei ($^{4}$He, C) the resonance
effect and the contribution of electromagnetic interactions are
much lower but there is no precision data on neutron scattering by
these nuclei up to now.

In most publications the formalism used for the extracting the
amplitudes of long-range interactions is described too shortly so
that many important details are omitted. We use here the method
close to proposed in \cite{Sam1} and most appropriate for analysis
of cross sections for high enriched samples, in which the
overwhelming contribution comes from one isotope. The experimental
values of coherent scattering lengths are not used in this
approach. The modification in our approach consists in the
following: we do not subtract the resonance contribution from
experimental data before fitting the experimental data but include
all resonance terms into the expression for the scattering
amplitude and fit the experimental data to calculated cross
section. Such procedure seems more reliable because resonance
contribution to the scattering amplitude interferes with potential
neutron-nucleus scattering amplitude and with the amplitudes of
the long-range interactions.

Scattering amplitude
\begin{equation}
f(q)=f_{0}(q)+f_{1}(q)+f_{\tiny{\mbox{Sch}}}(q)+f_{ne}(q)+f_{p}(q)+f_{w}(q),
\end{equation}
where $f_{0}(q)$ is the amplitude of nuclear $s$-scattering,
$f_{1}(q)$ is the amplitude of nuclear $p$-scattering,
$f_{\tiny{\mbox{Sch}}}$ is the amplitude of electromagnetic
(Schwinger) scattering, $f_{ne}$ is the amplitude of $n-e$
scattering, $f_{p}$ is the amplitude of neutron-nucleus scattering
due to neutron electric polarizability, $f_{w}$ is the amplitude
of hypothetical additional long-range neutron-nucleus interaction.

Transmission cross-section
\begin{eqnarray}
\sigma(k)=\int\nolimits|f(q)|^{2}d\Omega & \approx &
4\pi\Big|f_{0}(k)+f_{ne}(k)+f_{p}(k)+f_{w}(k)\Big|^{2}+
12\pi\Big|f_{1}(k)\Big|^{2}+ \nonumber\\
 & & +\sigma_{1p}+\sigma_{1w}+
\sigma_{a}+\sigma_{\tiny{\mbox{Sch}}}+\sigma_{\tiny{\mbox{inc}}}+
\Delta\sigma_{\tiny{\mbox{sol}}}.
\end{eqnarray}

In this expression $f_{i}(k)=\int\nolimits f_{i}(q)d\Omega/4\pi$,
where "i" stands for $"0","ne","p"$, and $"w"$;
$|f_{1}(k)|^{2}=\int\nolimits|f_{1}(q)|^{2}d\Omega/12\pi$; instead
of the terms $\int\nolimits|f_{i}|^{2}d\Omega$ are used the terms
$|\int\nolimits f_{i}d\Omega|^{2}$. The latter is valid for the
amplitude $f_{0}$ which does not depend on the scattering angle,
and for the amplitudes $f_{n-e}, f_{p}$, and $f_{w}$ which are
from three to four orders of magnitude lower than nuclear
scattering amplitude $f_{0}$. The error induced by such an
approximation as was shown by direct calculation does not exceed
the tenth of millibarn that is significantly below the precision
of the total neutron cross section by heavy neuclei, which in the
best case is not better than several millibarns. $\sigma_{a}$ is
the capture cross section, $\sigma_{\tiny{\mbox{Sch}}}$ is the
Schwinger scattering cross section due to the interaction of
neutron magnetic moment with nuclear electric charge,
$\Delta\sigma_{\tiny{\mbox{sol}}}$ is the correction to the
scattering cross section due to solid state structure and dynamics
effects, and $\sigma_{\tiny{\mbox{inc}}}$ is the incoherent cross
section \cite{inc}.

We do not neglect here small terms in the scattering cross section
due to interference of $p$-scattering with the amplitudes $f_{p}$,
($\sigma_{1p}$) and $f_{w}$ ($\sigma_{1w}$).

\begin{equation}
f_{0}=\frac{i}{2k}(1-S_{0}),
\end{equation}
$k=2.1968\times 10^{-4}\frac{A}{A+1}E^{1/2}$ is the neutron wave
vector (fm$^{-1}$) in center of mass system (neutron energy in the
lab. system in eV).

S-matrix
\begin{equation}
S_{0}=e^{2i\delta_{0}}\Biggl(1-\sum_{j}\frac{ig_{j}\Gamma_{nj}}
{(E-E_{j})+i\Gamma_{j}/2}\Biggr),
\end{equation}
where summation is performed over known $s$-resonances,
$\Gamma_{nj}=\Gamma_{nj}(E_{j})k/k_{j}$,
$\Gamma_{j}=\Gamma_{nj}+\Gamma_{\gamma j}$, and $g_{j}$ are the
neutron widths, the total widths, and the statistical weights of
$j^{th}$ $s$-resonance.

Upon introducing
\begin{equation}
\sum\nolimits_{(1)}=\sum_{j}\frac{2}{k_{j}}\frac{g_{j}\Gamma^{0}_{nj}
(E-E_{j})}{4(E-E_{j})^{2}+\Gamma_{j}^{2}},
\end{equation}
and
\begin{equation}
\sum\nolimits_{(2)}=\sum_{j}\frac{1}{k_{j}}\frac{g_{j}
\Gamma^{0}_{nj}\Gamma_{j}}{4(E-E_{j})^{2}+\Gamma_{j}^{2}},
\end{equation}
where $\Gamma^{0}_{nj}$ are the reduced neutron widths, the real
and the imaginary parts of the nuclear scattering amplitude are
\begin{equation}
{\mbox{Re}}f_{0}=\frac{{\mbox{sin}}(2\delta_{0})}{2k}-
{\mbox{cos}}(2\delta_{0})\sum\nolimits_{(1)}-
{\mbox{sin}}(2\delta_{0})\sum\nolimits_{(2)}
\end{equation} and
\begin{equation}
{\mbox{Im}}f_{0}=\frac{{\mbox{sin}}^{2}\delta_{0}}{k}+
{\mbox{cos}}(2\delta_{0})\sum\nolimits_{(2)}-{\mbox{sin}}(2\delta_{0})
\sum\nolimits_{(1)}.
\end{equation}

For the distant and unknown resonances, when $E\ll E_{j'}$, and
$\Gamma_{\gamma j'}\ll \Gamma_{nj'}\ll E_{j'}$, we use the
expansion
\begin{equation}
\sum\nolimits_{(i')}\approx\sum\nolimits_{(i')}(k=0)+
\sum\nolimits_{(i')}^{'}(k=0)k+\sum\nolimits_{(i')}^{''}(k=0)k^{2}/2,
\end{equation}
$(i'$=1,2) so that:
\begin{equation}
\sum\nolimits_{(1')}\approx-\sum_{j'}\frac{g_{j'}\Gamma_{nj'}}{2k_{j'}E_{j'}}+
\sum_{j'}\frac{g^{2}_{j'}\Gamma^{2}_{nj'}\Gamma_{\gamma j'}}
{4k^{2}_{j'}E^{3}_{j'}}k-
\sum_{j'}\frac{g_{j'}\Gamma_{nj'}}{k_{j'}E_{j'}^{2}}E,
\end{equation}
and
\begin{equation}
\sum\nolimits_{(2')}=\sum_{j'}\frac{g_{j'}\Gamma_{nj'}\Gamma_{\gamma j'}}
{4k_{j'}E^{2}_{j'}}+\sum_{j'}\frac{g^{2}_{j'}\Gamma^{2}_{nj'}}
{4k^{2}_{j'}E^{2}_{j'}}k+
\sum_{j'}\frac{g^{3}_{j'}\Gamma^{3}_{nj'}\Gamma_{\gamma j'}}
{8k_{j'}^{3}E_{j'}^{4}}k^{2}.
\end{equation}

The contribution of these distant resonance terms to the cross
section term proportional to $k$ is:
\begin{equation}
8\pi(R\sum\nolimits_{(1')}^{'}(0)-R^{2}\sum\nolimits_{(2')}(0)+
\sum\nolimits_{(1')}(0)\sum\nolimits_{(1')}^{'}(0)
+\sum\nolimits_{(2')}(0)\sum\nolimits_{(2')}^{'}(0)),
\end{equation}
and the contribution of these distant resonance terms to the cross
section term proportional to $k^{3}$ is:
\begin{eqnarray}
\frac{4\pi}{3}(2R^{4}\sum\nolimits_{(2')}(0)-3R^{2}\sum\nolimits_{(2')}^{''}
(0)-4R^{3}\sum\nolimits_{(1')}^{'}(0)+ \nonumber\\
+3\sum\nolimits_{(1')}^{'}(0)\sum\nolimits_{(1')}^{''}(0)
+3\sum\nolimits_{(2')}^{'}(0)\sum\nolimits_{(2')}^{''}(0)).
\end{eqnarray}
The calculation of these terms with sum over all known \cite{Sukh}
resonances for the sample with isotopic content
$^{208}$Pb$_{0.983}$\quad$^{207}$Pb$_{0.011}$\quad$^{206}$Pb$_{0.006}$
\cite{is3,is8} yields the contribution for $k$-proportional term
less than 10$^{-6}$ b, and for  $k^{3}$-proportional term
$\sim$10$^{-5}$ b at the neutron energy about 20 keV.

As at low energies $\delta=-kR^{'}_{n}$, where $R^{'}_{n}$ is the nuclear part
of the scattering radius, expansion of the first term in Re$f_{0}$ gives:
\begin{eqnarray} {\mbox{Re}}f_{0} & = &
-R^{'}_{n}+\frac{2}{3}R^{'3}_{n}k^{2}-
\frac{2}{15}R^{'5}_{n}k^{4}+\frac{4}{315}R^{'7}_{n}k^{6}-
 \frac{2}{2835}R^{'9}_{n}k^{8}- \nonumber\\ & &
-{\mbox{cos}}(2\delta_{0})\sum\nolimits_{(1)}-
{\mbox{sin}}(2\delta_{0})\sum\nolimits_{(2)}+h_{0}+h_{1}k^{2}.
\end{eqnarray}

The last two terms represent the contribution from the tails of
distant and unknown (including at negative energies) resonances,
and comes from the expansion of the term
cos(2$\delta_{0})\sum\nolimits_{(1')}$. At low energies $E\ll
|E_{j'}|$, and $\Gamma_{j'}\ll |E_{j'}|$ this contribution is:
\begin{equation}
-\sum_{j'}\frac{g_{j'}\Gamma_{nj'}}{2k_{j'}E_{j'}}-
\sum_{j'}\frac{g_{j'}\Gamma_{nj'}}{2k_{j'}E_{j'}^{2}}E=h_{0}+h_{1}k^{2}.
\end{equation}
The first term enters into unknown scattering radius $R'$, the second term we
have to consider as unknown (fitted) parameter.

The amplitude of the neutron-atom scattering due to the neutron-electron
interaction is
\begin{equation}
f_{ne}=-b_{ne}Z(\bar f-\bar h),
\end{equation}
where $b_{ne}$ is neutron-electron scattering length,
\begin{equation}
\bar f(k)=\frac{2}{x}(\sqrt{(1+x)}-1),\qquad
x=12(k/q_{0})^{2},\qquad q_{0} ({\mbox{fm}}^{-1})=1.9\times
10^{-5}Z^{1/3}
\end{equation}
is the atomic formfactor, and
\begin{equation}
\bar h(k)=1-\frac{(kR_{N})^{2}}{5}+2\frac{(kR_{N})^{4}}{105}-...
\end{equation}
is the nuclear charge fromfactor \cite{Sears}.

$R_{N}=1.2027 A^{1/3}=7.127$ fm for $^{208}$Pb is the nuclear electric radius.

The neutron electric polarizability scattering amplitude
\begin{equation}
f_{p}=f_{p0}\cdot Q,
\end{equation}
where
\begin{equation}
f_{p0}=\alpha_{n}\Bigl(\frac{Ze}{\hbar}\Bigr)^{2}\frac{m}{R_{N}}=32.78
\, {\mbox{mfm}}
\end{equation}
for Z=82 and the neutron electric polarizability
$\alpha_{n}=1.2\times 10^{-3}$ fm$^{3}$, the formfactor
\begin{equation}
Q=\frac{6}{5}-\frac{\pi}{3}\Bigl(kR_{N}\Bigr)+\frac{2}{7}
\Bigl(kR_{N}\Bigr)^{2}-\frac{4}{405}\Bigl(kR_{N}\Bigr)^{4}...
\end{equation}

The $p$-scattering amplitude is
\begin{equation}
f_{1}=\frac{i}{2k}(1-S_{1}),
\end{equation}
\begin{equation}
S_{1}=e^{2i\delta_{1}}\Biggl(1-\sum_{j}\frac{ig_{j}\Gamma_{nj}}
{(E-E_{j})+i\Gamma_{j}/2}\Biggr).
\end{equation}

Here summation is performed over known $p$-resonances,
$\Gamma_{j}=\Gamma_{nj}+\Gamma_{\gamma j}$, and $\Gamma_{j},\,
\Gamma_{nj},\, \Gamma_{\gamma j}$, and $g_{j}$ are the total
width, the neutron width, the gamma-width and the statistical
weight of $j^{th}$ $p$-resonance.
\begin{equation}
\Gamma_{nj}=\frac{k}{k_{j}}\Gamma_{nj}(E_{j})\frac{v_{1j}}{v_{0j}},
\end{equation}
where
\begin{equation}
v_{0j}=\frac{(k_{j}R)^{2}}{1+(k_{j}R)^{2}},\qquad
v_{1j}=\frac{(kR)^{2}}{1+(kR)^{2}}.
\end{equation}
The phase of $p$-scattering
\begin{equation}
\delta_{1}=-kR+{\mbox{arctg}}(kR)+
{\mbox{arcsin}}\Bigl[\frac{k}{3}\frac{(kR)^{2}}{1+(kR)^{2}}
\bigl(R-R^{'}_{1}\bigr)\Bigr],
\end{equation}
where $R$ is the channel radius, $R^{'}_{1}$ is the $p$-wave scattering
radius.

Interference terms of the $p$-scattering amplitude and the polarization
amplitude are:
\begin{equation}
\sigma_{1p}=4\pi {\mbox{Re}}f_{1}(k)f_{p}\Bigl(\frac{4\pi}{5}x-
\frac{2}{3}x^{2}\Bigr),
\end{equation}
where $x=kR$.

\section{Obtaining of constraints for hypothetical long-range interactions
from neutron scattering experiment.}
 The method to search for the admixture of long-range interaction
in the neutron-nucleus scattering was proposed by Thaler \cite{Th}
in application to the measurement of the neutron electric
polarizability.  The idea consists in the search for non-even in
the neutron wave vector terms in the power series of the
experimental angular distribution or the total neutron cross
sections in keV energy range.

Indeed, in result of interference of the Born scattering
amplitudes for the long-range potentials of the Eq.(4) and the
short range nuclear scattering amplitude, the following terms
appear in the low energy wave vector $k$ expansion for the total
cross section (Eq.(6)):
\begin{eqnarray}
f_{1w}=\frac{\alpha_{1}\pi br_{0}}{2k}\quad\mbox{ for }N=1, \qquad
f_{2w}=-\frac{1}{2}\alpha_{2}br_{0}^{2}{\mbox{ln}}(kR)\quad\mbox{ for }N=2,
\nonumber\\
f_{3w}=-\frac{1}{3}\alpha_{3}b\pi r_{0}^{3}k\quad\mbox{ for }N=3, \qquad
f_{4w}=\frac{1}{3R}\alpha_{4}br_{0}^{4}k^{2}{\mbox{ln}}(kR)\mbox{ for }N=4,
\nonumber\\
f_{5w}=\frac{1}{15}\alpha_{5}\pi br_{0}^{5}k^{3}\quad\mbox{ for }N=5, \qquad
f_{6w}=-\frac{2}{45}\alpha_{6}br_{0}^{6}{\mbox{ln}}(kR)k^{4}\mbox{ for }N=6.
\end{eqnarray}
Here $b=\frac{2m^{2}MG}{\hbar^{2}}$, $b=1.3\times 10^{-35}$
fm$^{-1}$ for the nucleus $^{208}$Pb.

The potentials for gravitating balls with radius $R$ for r$>R$ may
be obtained by integration of the potentials of the Eq.(4) over
the ball volume and have the form:
\begin{equation} N=1,   \quad
U_{1}(r)=\frac{3g_{1}r_{0}}{2R^{3}r}\Biggl[\frac{R^{2}-r^{2}}{2}{\mbox{ln}}
\Bigl|\frac{R+r}{R-r}\Bigr|+Rr\Biggr];
\end{equation}
\begin{equation}
N=2,   \quad
U_{2}(r)=\frac{3g_{2}r_{0}^{2}}{R^{3}r}\Biggl[\frac{r}{2}{\mbox{ln}}\Bigl|
\frac{R+r}{R-r}\Bigr|-R\Biggr];
\end{equation}
\begin{equation}
N=3,   \quad
U_{3}(r)=\frac{3g_{3}r_{0}^{3}}{2R^{3}}\Biggl[-\frac{1}{2r}{\mbox{ln}}\Bigl|
\frac{R+r}{R-r}\Bigr|+\frac{R}{r^{2}-R^{2}}\Biggr];
\end{equation}
\begin{equation}
N=4,   \quad U_{4}(r)=\frac{g_{4}r_{0}^{4}}{r(r^{2}-R^{2})^{2}};
\end{equation}
\begin{equation}
N=5,   \quad U_{5}(r)=\frac{g_{5}r_{0}^{5}}{(r^{2}-R^{2})^{3}};
\end{equation}
\begin{equation}
N=6,   \quad
U_{6}(r)=\frac{g_{6}r_{0}^{6}}{5r}\frac{R^{2}+5r^{2}}{(r^{2}-R^{2})^{4}}.
\end{equation}

Born scattering amplitudes for these potentials
\begin{equation}
f_{N}=b\frac{\alpha_{N}}{q}\int \limits_{R+\epsilon}^{\infty}V_{N}(r)
{\mbox{sin}}(qr) r dr,
\end{equation}
where $V_{N}=U_{N}/g_{N}$, \,  $\epsilon$ is the cut-off parameter
in extra-gravity interaction, it is taken at the level of
electroweak scale $\sim$10$^{-4}$ fm \cite{Ark}. These amplitudes
have very complicated view, we are interested in the low neutron
energy expansions of these amplitudes.

For example, if to omit in these expansions the constant and even in the wave
vector terms (which is impossible to distinquish from contribution from
short-range nuclear interaction) we have the following contributions to
the formfactor in the total scattering cross section:
\begin{equation}
f_{4w}=\frac{1}{6}\alpha_{4} b r_{0}^{4}\Bigl[{\mbox{ln}}(kR)+
{\mbox{ln}}(k\epsilon)\Bigr]\cdot k^{2}
\end{equation}
for the potential of Eq.(36),
\begin{equation}
f_{5w}=\frac{\pi}{15}\alpha_{5} b r_{0}^{5}
\Bigl[1+\frac{2\pi}{7}(kR)^{2}\Bigr]\cdot k^{3}
\end{equation}
for the potential of Eq.(37) and
\begin{equation}
f_{6w}=\frac{1}{45}\alpha_{6} b r_{0}^{6}\Bigl[{\mbox{ln}}(kR)+
{\mbox{ln}}(k\epsilon)\Bigr]\cdot k^{4}
\end{equation}
for the potential of Eq.(38). The first terms in the brackets
correspond to the pointlike potentials of Eqs.(4), the remaining
terms come from the fact that the sources of potentials are
spherical bodies. For the potentials of Eqs.(33-35) with N$\leq$3
no additional terms appear in these expansions for the scattering
amplitudes for gravitating balls in comparison to the cases of the
pointlike scatterers of Eqs.(32).

As an example of the measurement of the total neutron cross
section in eV--keV energy range we used two sets of data from the
publications \cite{is3,is8} and \cite{Gat}.  These groups measured
the total cross section of highly enriched $^{208}$Pb samples with
atomic concentations: $^{208}$Pb$_{0.983}\, ^{207}$Pb$_{0.011}\,
^{206}$Pb$_{0.006}$ in the energy range from 1.26 eV up to 24 keV
with declared precision from $\sim$ 20 mb to 2.5 mb in
\cite{is3,is8} and $^{208}$Pb$_{0.9712}\, ^{207}$Pb$_{0.0179}\,
^{206}$Pb$_{0.0108}$ in \cite{Gat} with similar precision.

The outlined formalism was used for the procedure of extracting
the contribution of possible new long-range interactions to the
total neutron cross section.  The MINUIT program \cite{min} was
used with three fitted parameters: the $s$-scattering radius
$R'_{n}$, the term corresponding to the contribution to the
$s$-scattering amplitude from distant and unknown resonances
$h_{1}$, and contribution from the unknown long-range potential
$f_{w}$.  The constraints obtained from both sets of data:
\cite{is3,is8} and \cite{Gat} were very close.

The neutron-electron scattering length was varied between
$b_{ne}=-1.32\times10^{-3}$ fm and $b_{ne}=-1.59\times 10^{-3}$
fm[28--32], the value of the neutron electric polarizability was
varied in the limits $\pm$30\% around $\alpha_{n}=(1.22\pm
0.19)\times 10^{-3}$ (fm)$^{3}$, the latter value being obtained
as an averaged over the results from two recent measurements of
the Compton scattering by deuteron \cite{pol1,pol2}.

Figure 1 shows obtained limits for non-Newtonian gravity from these neutron
scattering data in terms of relative potential strength $\alpha$ in
Eq.(1) as a function of the Yukawa length scale $\lambda$.
It shows also the present status of constraints from the previous larger
scales experiments.

Figure 2 shows the limits on the value of the inverse power law
potential of Eq.(4), parametried in the form:
\begin{equation}
U(r)=U_{0} (1 {\mbox{fm}})M\Bigl(\frac{1
{\mbox{fm}}}{r}\Bigr)^{N+1},
\end{equation}
where $M$ is the relative nuclear mass.

It is hardly believable that radical increase in precision of
keV-neutron scattering cross section may be achieved in the
forthcoming years. Analysis of systematic effects and
uncertainties in the processing data of neutron scattering arising
mostly from resonance structure of cross section shows that the
limiting precision lies in the region 5 mb. Not the least is the
problem of correct accounting the background in the time of flight
measurements \cite{is3,is8,Gat,Ar}:  different methods of
measurement and subtracting the background yield significantly
differing results.

Possibly some increase in sensitivity in the search for new
interactions in the neutron-nuclei scattering may be achieved by
measuring the energy dependence of the asymmetry of slow neutron
scattering by noble gases - the method of Fermi--Marshall
\cite{F-M,Kro}.

The neutron-atom scattering amplitude
\begin{equation}
a(\theta)=a_{N}+a_{n-e}Zf(\theta)
\end{equation}
in result of interference of the nuclear and electron scattering leads to the
scattering cross section;
\begin{equation}
\frac{d\sigma(\theta)}{d\Omega}=a_{N}^{2}+2a_{N}a_{n-e}Zf(\theta),
\end{equation}
with the relative value of the anisotropic term
\begin{equation}
\delta\Bigl(\frac{d\sigma(\theta)}{d\Omega}\Bigr)=2\frac
{a_{n-e}}{a_{N}}Zf(\theta).
\end{equation}
Atomic formfactor may be calculated with sufficient precision according
to \cite{Sears}
\begin{equation}
f(q)=\Bigl[1+3\bigl(\frac{q}{q_{0}}\bigr)^{2}\Bigr]^{-1/2},
\end{equation}
with
\begin{equation}
q_{0}=\beta Z^{1/3}(\AA^{-1}).
\end{equation}
In these expressions $a_{N}$ is the nuclear scattering amplitude,
$a_{n-e}$ is the amplitude of n-e scattering, Z is the charge
number. The value of $\beta$ is taken from the tables
\cite{X-Tab}.

Significant diffraction effect appears in the neutron scattering by noble
gases due to atom--atom interactions[38--43].
The static structure factor depends on interatomic interaction potential
$U(r)$ and atomic density $n$:
\begin{equation}
S(q)=1+n\int \limits_{0}^{\infty}\Bigl(e^{-U(r)/kT}-1\Bigr)
e^{i\mbox{\bf qr}}d^{3}\mbox{\bf r}.
\end{equation}
For the spherically symmetric potential $U(r)$ the structure factor is:
\begin{equation}
S(q)=1+n\frac{4\pi}{q}\int
\limits_{0}^{\infty}\Bigl(e^{-U(r)/kT}-1\Bigr) {\mbox{sin}}(qr)r
dr.
\end{equation}
The scattering asymmetry is introduced as the ratio of intensities of the
scattered neutrons to the angles $\theta_{1}$ and $\theta_{2}$, (usually
$\theta_{1}+ \theta_{2}=\pi$).
\begin{equation}
\frac{S(\theta_{1})}{S(\theta_{2})}.
\end{equation}
Figure 3 shows the results of the calculations of the effects of hypothetical
interactions on the center-of-mass scattering asymmetry by Xe gas together
with the effects of neutron diffraction and neutron-electron interaction.
The curves show the results of calculation of asymmetry for the forward and
backward angles, respectively, 30$^{\circ}$ and 150$^{\circ}$.  For Xe several
best approximations for the potential of interatomic interaction were taken
from the literature.

\section{Constraints for hypothetical long-range interactions from
antiprotonic atoms.}
 Some constraints on long-range forces may be
obtained from the measurements of energies of electric-dipole
transitions in hadronic atoms. The average distance of hadron from
nucleus is
\begin{equation}
<r>\simeq\frac{200}{Z}\frac{m_{\pi}}{m_{r}}\frac{3n^{2}-l(l-1)}{2}
{\mbox{(fm)}},
\end{equation}
where $n$ and $l$ are the quantum numbers of the state, and
$m_{r}$ is the reduced mass of the bound hadron. This distance is
in the range of dozens to thousand of fermies for studied
antiprotonic atoms. The most convenient transitions between the
states ($n,l$) are of the type ($n,n-1)\rightarrow(n-1,n-2$) with
$n$ large enough to decrease as much as possible the effects of
strong interaction on the energy states of orbiting hadrons. To
estimate the contribution to the energy shifts of these states
from the potentials of Eq.(1) and Eq.(4) we have to calculate the
expectation value in the hydrogenlike atomic states of these
potentials: They are
\begin{equation}
\Delta E_{n}=\frac{gZ}{n^{2}a_{0}} \Bigl(\frac{2\lambda Z}{2\lambda Z+n
a_{0}}\Bigr)^{2n},
\end{equation}
where $Z$ is nuclear charge, $a_{0}=\hbar^{2}/\mu e^{2}$, for the potential of
Eq. (1) and
\begin{equation}
\Delta E_{n}=\frac{g_{N}}{r_{0}}\Bigl(\frac{2Zr_{0}}{na_{0}}\Bigr)^{(N+1)}
\frac{(2n-N-1)!}{(2n)!}
\end{equation}
for potentials of Eq.(4). The difference of experimental energies
of transitions $E(n,l)-E(n-1,l-1)$ and the calculation of the QED
predictions for these transition energies may be assumed to be the
effect of long-range interaction. No real difference was observed
in the most precise experiments we used for obtaining these
restrictions. We used the reported values of experimental
uncertainty 4$\times $10$^{-7}$ eV for transitions
(32,31)$\rightarrow$(31,30) in antiprotonic helium \cite{Hory} and
the value 50 eV for transitions (5,4)$\rightarrow$(4,3) in
antiprotonic sulfur \cite{Bamb} as the boundaries for the
difference in the energy shifts due to these long-range
interactions. Figure 1 shows these constraints as a function of
$\lambda$ for the potential of Eq.(1). For the potential of Eq.(4)
similar constraints may be obtained assuming that the
compactification radius R$_{\mbox{comp}}\gg<r>$ -- the average
distance of the orbiting antiproton from the nucleus. The latter
is calculated according to Eq.(51) and is $\sim$ 0.2\,\AA\enspace
for the case of antiprotonic helium \cite{Hory}, and $\sim$ 50 fm
for the case of antiprotonic sulfur \cite{Bamb}. The constraints
for the potential of Eq.(4) are $\alpha_{2}\geq 6\times$10$^{33}$
for the measurement \cite{Bamb} and $\alpha_{2}\geq
6\times$10$^{32}$ for the measurement \cite{Hory}.

The possibility of probing additional space dimensions with fast
neutron scattering experiments was considered recently in
\cite{Mex}. I am grateful to Dr. W. M. Snow (Indiana University)
for attracting my attention to this paper. The author is grateful
also to Dr. A. B. Popov for important comment and to Dr. G. S.
Samosvat for consultations and discussions.

A. B. Laptev, Yu. A. Alexandrov, I. S. Guseva, {\it et al.}, in
{\it the Proceedings of X Intenational Seminar on Interaction of
Neutrons with Nuclei, Dubna, 22--25 May 2003}, p.79.

\bibitem{min}
F. James, CERN Program Library Long Writeup D506, Reference Manual, CERN,
1994.

\bibitem{Mel}
E. Melkonian, B. M. Rustad, and W. W. Havens, Phys. Rev. {\bf 114}, 1571
(1959).

\bibitem{Alex}
Yu. A. Alexandrov, T. A. Machekhina, L. N. Sedlakova and L. E.
Fykin, Yad. Fiz. {\bf 60}, 1190 (1974); [Sov. J. Nucl. Phys. {\bf
60}, 623 (1975)]; Yu. A. Alexandrov, Z. Phys. A{\bf 344}, 219
(1992).

\bibitem{Kro}
V. E. Krohn and G. Ringo,  Phys. Rev. {\bf 148}, 1303 (1966);
Phys. Rev. D{\bf 8}, 1305 (1973).

\bibitem{Koe}
L. Koester, W. Washkowski, and A. Kluver, Physica. {\bf 137}B, 282 (1986).

\bibitem{Kop}
S. Kopecky, J. A. Harvey and N. W. Hill, {\it et al.}, Phys. Rev.
C{\bf 56}, 2229 (1997).

\bibitem{pol1}
K. Kossert, M. Camen, F. Wissmann, {\it et al.}, Phys. Rev. Lett.
{\bf 88}, 162301-1 (2002).

\bibitem{pol2}
M. Lundin, J.-O. Adler, M. Boland, {\it et al.}, Phys. Rev. Lett.
{\bf 90}, 192501-1 (2003).

\bibitem{Ar}
T. L. Enik, V. A. Ermakov, R. V. Kharjuzov, {\it et al}., Yad. Fiz.  {\bf 66},
59 (2003).

\bibitem{F-M}
E. Fermi and L. Marshall, Phys. Rev. {\bf 72}, 1139 (1947).

\bibitem{X-Tab}
{\it The International Tables for X-Ray Crystallography}, Ed. by
J. A. Ibers and W. C. Hamilton (Kynoch Press, Birmingham, 1974),
Vol.IV.

\bibitem{LJ-Xe1}
J. J. H. Van den Biesen, F. A. Stokvis, E. H. van Veen and C. J.
N. van den Meijdenberg, Physica {\bf 100}A, 375 (1980).

\bibitem{LJ-Xe2}
G. C. Maitland and E. B. Smith, {\it Seventh Symposium on
Thermophysical Properties}, Ed. by A. Cezairlian (Am. Soc. Mech.
Eng., 1977).

\bibitem{BWLSL-Kr-Xe}
J. A. Barker, R. O. Watts, J. K. Lee, {\it et al}., Chem. Phys. {\bf 61}
3081 (1974).

\bibitem{And-Ar}
C. D. Andriesse and E. Legrand, Physica {\bf 57}, 191 (1972).

\bibitem{Fre-Ar}
H. Fredrikze, C. D. Andriesse, and E. Legrand E., Physica {\bf 62}, 474
(1972).

\bibitem{Fred-Ar}
H. Fredrikze, J. B. van Tricht, Ad. D. van Well,  {\it et al.},
Phys. Rev. Lett. {\bf 62}, 2612 (1989).

\bibitem{Mex}
A. Frank, P. Van Isacker, and J. Gomez-Camacho, Phys. Lett. {\bf B 582}, 15
(2004).

\newpage
\listoffigures
Fig. 1.  Bounds $\alpha_{G}$ vs. $\lambda$ (90\% C.L.) from different
non-Newtonian gravity experiments. The area above the lines is excluded:
{\it 1} -- Cavendish--type experiment of Seattle group\cite{Seattle};
{\it 2} -- microcantilever Cavendish--type experiment of Stanford\cite{Stanf};
{\it 3} -- Casimir force measurements; constraints from\cite{cas1,Rep}, based
on earlier experiments of\cite{Der,Isra};
{\it 4} -- Casimir force measurement\cite{Lam}, the bound from\cite{Ab};
{\it 5} -- Casimir force measurements with atomic force microscope\cite{Moh},
bounds from\cite{Rep};
{\it 6} -- neutron quantum levels in the Earth gravitational field,
experiment\cite{nesv}, bounds from\cite{Ab};
{\it 7} -- van der Waals force measurements\cite{Isra}, bounds
from\cite{cas2,Rep};
{\it 8} -- atomic-force microscopy experiment\cite{Moi};
{\it 9} -- bounds obtained in this work from $X$-ray energy measurements in
antiprotonic sulfur atoms\cite{Bamb};
{\it 10} -- bounds obtained in this work from the measurements of
the transition frequencies between large $n$ levels in antiprotonic
He-atoms\cite{Hory};
{\it 11} -- bounds obtained in this work from analisys of
neutron total cross section scattering by $^{208}$Pb nuclei\cite{is3,is8};

\vspace{5mm}

Fig.2. Limits (90\% C.L.) on the value of the interaction potential
U$_{0}(1fm)$ vs $N$ in the expression
U(r)=U$_{0}$(1fm)M$\Bigl(\frac{1fm}{r}\Bigr)^{N+1}$:
{\it 1} -- from $X$-rays in $\tilde p-^{4}$He - atoms;
{\it 2} -- from $X$-rays in $\tilde p-$S -- atoms;
{\it 3} -- from the measurement of the total $n-^{208}$Pb -- scattering;
{\it 4} - the same as 3, but additional interference terms for scatterting
amplitudes for gravitating balls of Eqs. (40-42) were included to the
expression for the total scattering cross section of Eq.(6).

\vspace{5mm}

Fig. 3. The calculated neutron energy dependence of the center-of-mass
scattering asymmetry ($\theta_{1}=\pi/6, \theta_{2}=5\pi/6$) from 1 atm
pressure Xe gas target in result of different effects:
{\it 1} and 2 -- $n--e$ interaction with the values of $n--e$ scattering
lengths -1.32$\cdot$10$^{-3}$fm and -1.59$\cdot$10$^{-3}$fm;
{\it 3, 4, 5}, and {\it 6} -- diffraction effect for different approximations
of the interatomic Xe--Xe interaction:
{\it 3} and {\it 4} - two approximations of Lennard-Jones potential for
Xe\cite{LJ-Xe1,LJ-Xe2},
{\it 5} and {\it 6} - potentials of more complicated form
from\cite{BWLSL-Kr-Xe};
{\it 7, 8, 9} and {\it 10} - effect of hypothetical potential of
the form Eq.(1) for $\lambda$ equal to 10$^{5}$fm, 10$^{4}$fm, 10$^{3}$fm and
10$^{2}$fm and $\alpha$ equal to 10$^{23}$, 10$^{25}$ , 10$^{27}$ and
10$^{29}$ respectively;
{\it 11, 12} and {\it 13} - effect of hypothetical potential of
the form Eq.(4) for $N=1$ and compactification radius 10$^{5}$fm, 10$^{4}$fm
and 10$^{4}$fm and $\alpha_{1}$ equal 10$^{29}$, 10$^{30}$ and 10$^{31}$
respectively.

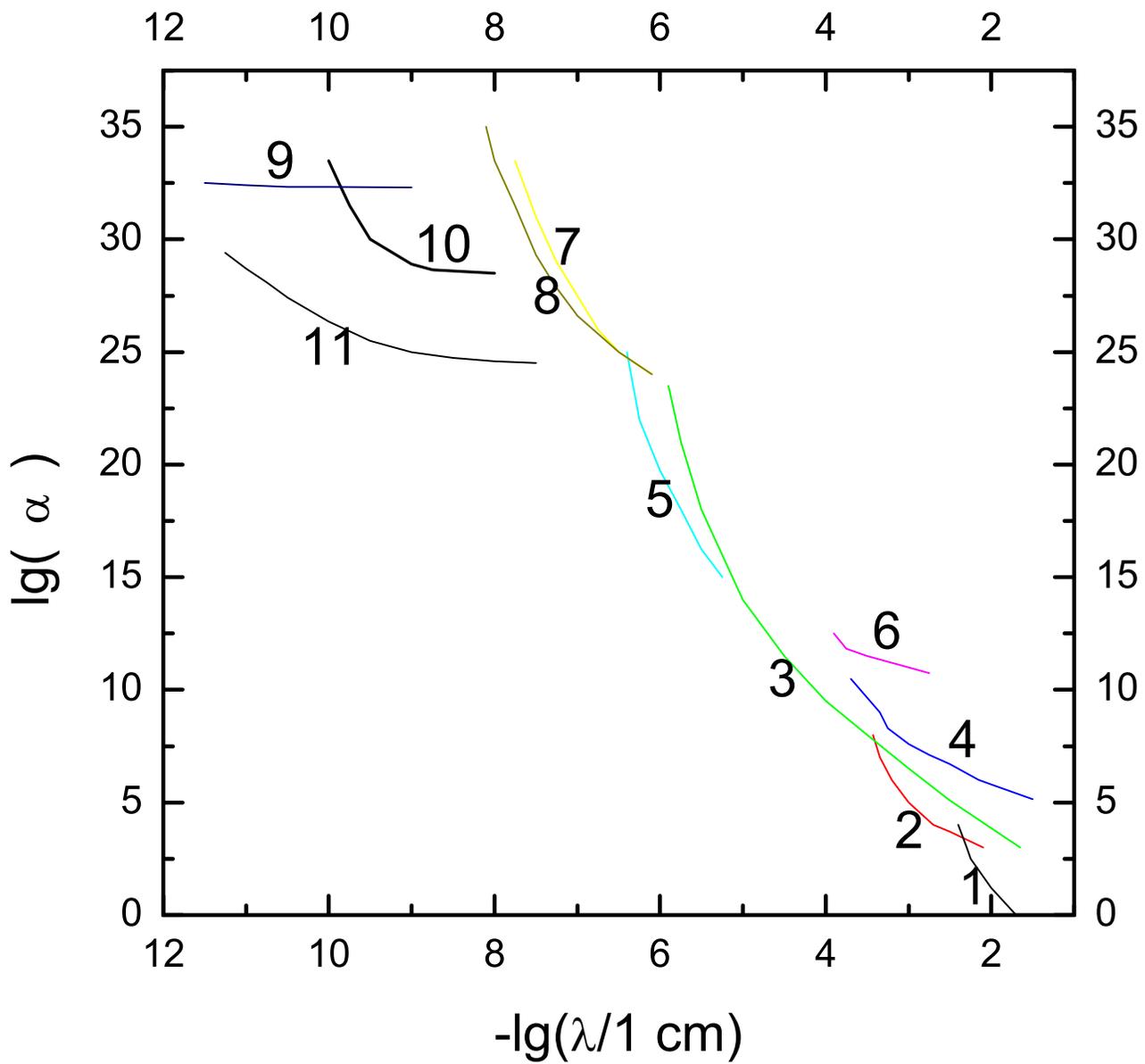
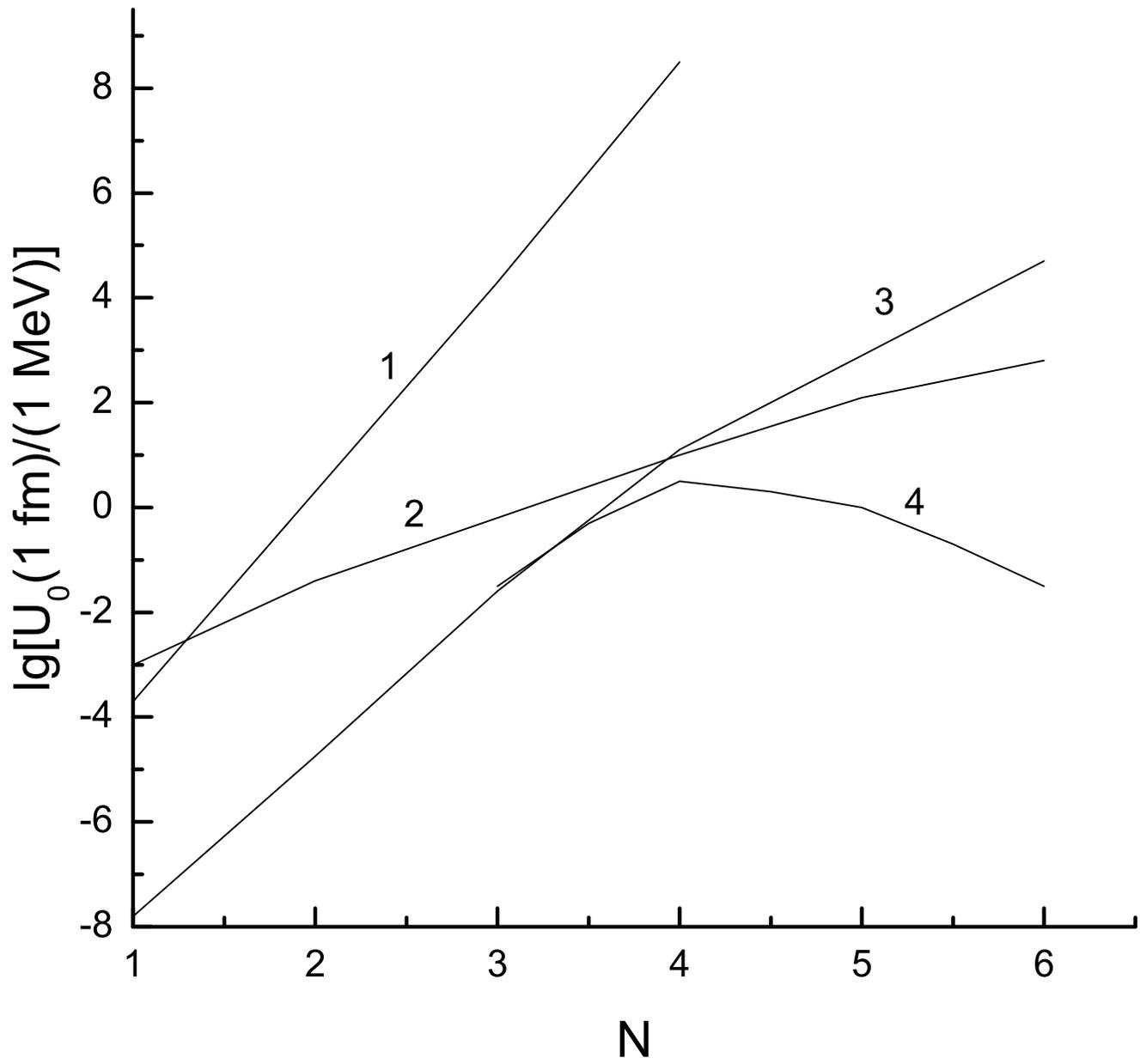
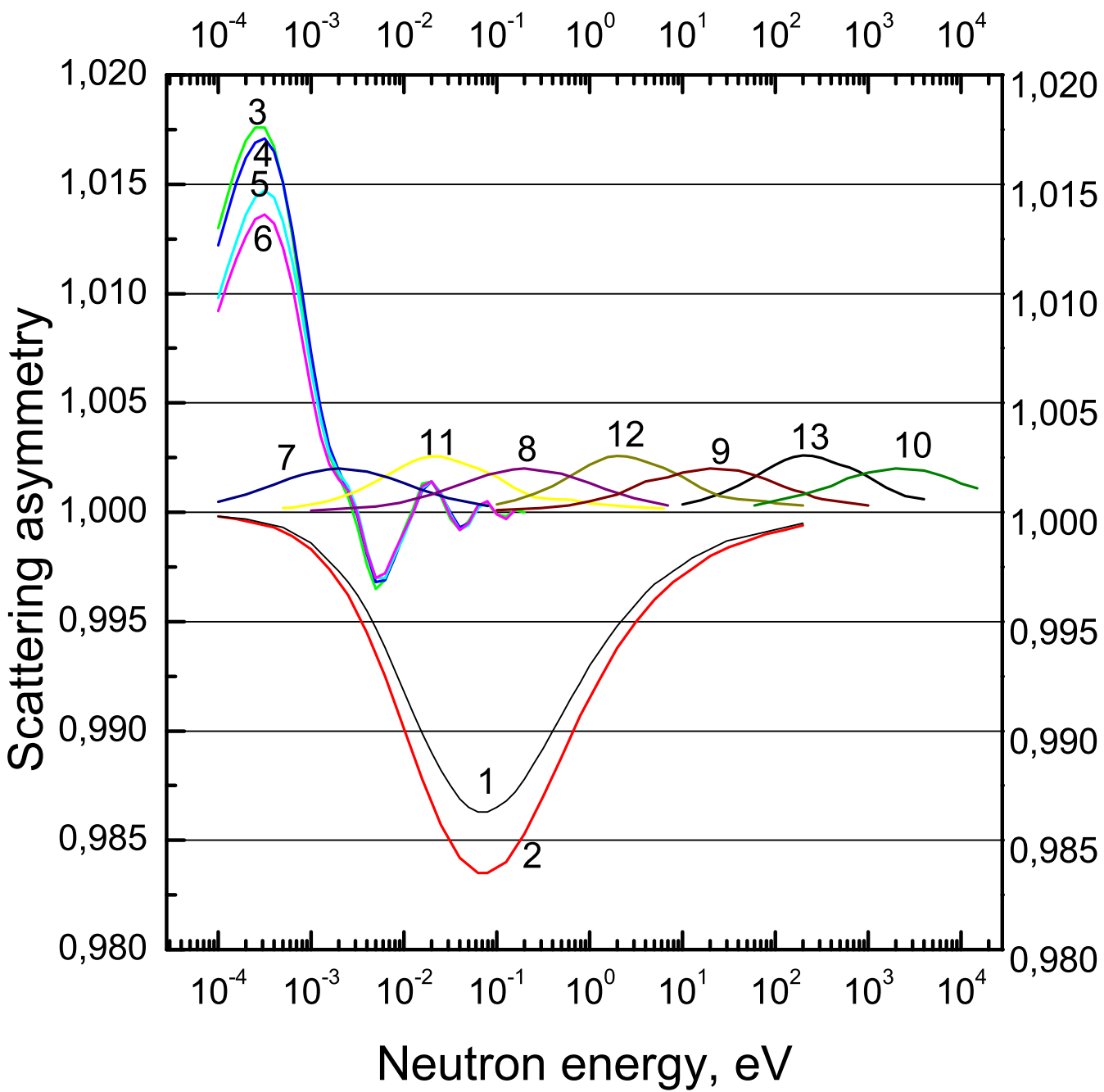
\end{document}